\documentclass[twocolumn,floatfix,groupedaddress,superscriptaddress,prb,showpacs]{revtex4}

\usepackage{graphicx}
\usepackage{amsmath}
\usepackage{amssymb}
\usepackage{amsfonts}
\usepackage{dcolumn}
\usepackage{bbm}
\usepackage{float}
\usepackage{color}

\def\be{\begin{equation}}       \def\ee{\end{equation}}
\def\bea{\begin{eqnarray}}      \def\eea{\end{eqnarray}}
\def\bes{\begin{subequations}}  \def\ees{\end{subequations}}

\begin{document}
\title{One-dimensional fermionic systems after interaction quenches and their 
description by bosonic field theories}

\author{Simone A. Hamerla}
\email{hamerla@fkt.physik.tu-dortmund.de}
\affiliation{Lehrstuhl f\"{u}r Theoretische Physik I, 
Technische Universit\"{a}t Dortmund,
 Otto-Hahn Stra\ss{}e 4, 44221 Dortmund, Germany}

\author{G\"otz S. Uhrig}
\email{goetz.uhrig@tu-dortmund.de}
\affiliation{Lehrstuhl f\"{u}r Theoretische Physik I, 
Technische Universit\"{a}t Dortmund,
 Otto-Hahn Stra\ss{}e 4, 44221 Dortmund, Germany}

\date{\textrm{\today}}

\begin{abstract} 
We study the time evolution of two fermionic one-dimensional models (spinless fermions with nearest-neighbor
repulsion and the Hubbard model) 
exposed to an interaction quench for short and moderate times. 
The method used to calculate the time dependence is a semi-numerical approach based on the Heisenberg equation of motion. We compare the results of this
approach to results obtained by bosonization implying power law behavior.
Indeed, we find that power laws describe our results well, but our results raise
the issue which exponents occur.
For spinless fermions it seems that the Tomonaga-Luttinger parameters work well which
also describe the equilibrium low-energy physics.
But for the Hubbard model this is not the case. Instead, we
find that exponents from the bosonization around the initial
state work well. Finally, we discuss what can be expected for the
long-time behavior.
\end{abstract}


\maketitle


\section{Introduction}

\label{chap:intro}

Lately, the interest in nonequilibrium physics has risen significantly. This is due to the progress on
the experimental side where many experimental setups have been developed which can be used to study 
many-body systems far from equilibrium. 
Amongst these setups are femtosecond photospectroscopy \cite{perf06} and cold atomic gases trapped in optical
lattices \cite{greiner02,greiner02b}. 
Due to their excellent decoupling of the atoms from any environment the time evolution of atoms in 
optical lattices can be viewed as realizations of closed quantum systems.
The relevant issues comprise the temporal evolution on short, on intermediate and on long time scales.
In the present article, we focus on short and intermediate time scales.

Since these systems usually are in highly excited states involving many degrees of freedom 
developing quickly already on short time scales, 
a theoretical description is very challenging and requires new theoretical approaches.
In the last years,  techniques have been developed to deal with systems out of equilibrium. Amongst these are 
DMFT techniques \cite{schmi02,freer06,eck09}, time dependent DMRG \cite{daley04,white04}, light-cone renormalization \cite{enss12}, methods based on
CUT techniques \cite{moe08}, on variational approaches including mean-field theories 
\cite{schiro10,sciol11,gamba11},  on perturbative renormalization of Keldysh Green
functions \cite{mitra11}, and on QMC techniques  \cite{goth12}, for a review see Ref.\ \onlinecite{polk11}.

There are several ways to realize states far from equilibrium. A widely considered scenario are quenches, 
i.e., sudden changes in the intrinsic system parameters \cite{barth08}. In this work we focus on interaction quenches, where the interaction $H_{\text{int}}$ is abruptly changed  \cite{cazal06,kolla07,manmana07,moe08,uhr09,schiro10,dora11,mitra11,goth12}. 
We will focus on quenches where the interaction is switched from zero to a finite value.
Thus, the system is prepared initially in the ground state 
of a non-interacting Hamiltonian $H_0$. But the time evolution is governed by the interacting Hamiltonian 
$H=H_0+H_{\text{int}}$. In this way, the system is in a highly excited state with respect to $H$.

In the present work, we study the time evolution of two generic fermionic lattice models in one dimension (1D). 
Of course, bosonic models are investigated in other studies 
as well\cite{kolla07,sabio10,fiore10,iucci10,grand10,sciol10,sciol11,gamba11}.
One-dimensional models play a special role in the context of thermalization because in these models an exhaustive
number of conserved quantities may exist. Naturally, the existence of conserved quantities influences the dynamics of the system strongly \cite{rigol07}. 
The first model to be studied are spinless fermions with nearest-neighbor repulsion. 
The quench dynamics of this system is contrasted to the quench dynamics of the second model,  the one-dimensional Hubbard model which includes the spin degree of freedom.
Both models are integrable.

In equilibrium, one-dimensional models are tackled by many analytical and numerical methods
whose review is far beyond the scope of the present article.
For the purposes of the present article, it is sufficient to note that gapless one-dimensional models
with linear dispersion at low energies can be efficiently described by bosonic field theories
\cite{luthe75,halda80,meden92,penc93,voit95,miran03,giama04}.
Often, the bosonic fields can be taken to be without interactions at low energies.
For instance, this is the case for spinless fermions. For models including spin,
sine-Gordon models include the leading bosonic interaction \cite{voit95,giama04}.
Naturally, the question arises whether the same or similar bosonic field theories are also able to 
describe the nonequilibrium dynamics \cite{karra12b,coira12,pollm13}. 
Quenches of the non-interacting bosonic models are fairly well understood by now 
\cite{cazal06,uhr09,iucci09,rentr12}. A set of results for the sine-Gordon model is also available
\cite{sabio10,fiore10,iucci10,grand10,mitra11,mitra12}.

The quenches considered in this work start from non-interacting Fermi seas. For these initial states 
the momentum distribution displays a jump at the Fermi points at $\pm k_\text{F}$, 
where the occupation drops from unity to zero on passing from  $|k|< k_\text{F}$
to $|k|> k_\text{F}$.
Exposed to the quench, the jump in the momentum distribution of the system evolves in time
which provides a sensitive probe for the quench dynamics.

We calculate the time evolution of the jump in the momentum distribution 
by a semi-numerical approach based on the Heisenberg equations of motion. 
The method is used to investigate the behavior of the two models on short and intermediate times after the quench.
Here we focus on the non-oscillatory, smooth decay of the jump;  
thus we study not too strong quenches which remain in the metallic regime.
The oscillatory behavior occurring for strong quenches into insulating phases
or close to them was studied separately \cite{hamer13a} for the sake of clarity \cite{note}.
For the accessible times, the dynamics of the jump can be described by 
the non-interacting bosonic field theories. 
But we find evidence that the underlying parameters differ from their values at equilibrium.
We discuss various scenarios for the behavior at longer times.

The paper is organized as follows. In the next section, the models under study are presented and the
semi-numerical method used to calculate the time dependence of the jump is explained, 
and the expectations from bosonic field theories are recalled. In section III, results for the spinless fermion model and for the Hubbard model are shown and compared to the
field theoretic expectations. Section IV discusses the long-time behavior.
In the last section, the results are summarized.

%
\section{Models and Methods}
\label{chap:model}

\subsection{Models}
\label{ssec:models}

The first model under study are spinless fermions
\begin{align}
\label{eq:NN}
H_\text{NN} = -J \sum_{\langle i,j\rangle} 
\big(\hat{c}_{i}^\dagger\hat{c}_{j}^{\phantom\dagger} + \text{h.c.}\big)
+U(t)\sum_i \hat n_i \hat n_{i+1}
\end{align}
with nearest-neighbor repulsion (NN).
The operator $\hat{c}_{j}^\dagger$ ($\hat{c}_{j}^{\phantom\dagger}$) creates (annihilates) a particle
at site $j$ and $\hat n_j=\hat{c}_{j}^\dagger \hat{c}_{j}^{\phantom\dagger}$.
By a Jordan-Wigner transformation it can be mapped to
an anisotropic spin $S=1/2$ XXZ chain \cite{fradk91}. It is integrable by Bethe ansatz
\cite{yang66aa,yang66a}.

The second model is the Hubbard model \cite{hubba63,gutzw63,kanam63} 
comprising spin $\sigma=\uparrow, \downarrow$
\begin{align}
\label{eq:Hu}
H_\text{Hu} = -J \sum_{\langle i,j;\sigma\rangle} 
\big(\hat{c}_{i,\sigma}^\dagger\hat{c}_{j,\sigma}^{\phantom\dagger} + \text{h.c.}\big)
+U(t)\sum_i \hat n_{i,\uparrow} \hat n_{i,\downarrow}
\end{align}
which is governed by the hopping $J$ and the local repulsion $U$.
This model as well is exactly solvable by Bethe ansatz \cite{lieb68,essle05}.

We focus on quenches where the interaction is switched on $U(t)=\Theta(t)U\ge 0$ abruptly
and consider Fermi seas as initial states, i.e., the ground states for $U=0$.
Throughout, the band width $W=4J$ is used as natural
energy scale and consequently time is measured in the inverse band width $1/W$
since we set $\hbar$ to unity.

\subsection{Methods}
\label{ssec:methods}

  Below some quantities with spin $\sigma$ are denoted for the Hubbard model. 
  In the corresponding quantities without spin, i.e., for the spinless fermion model,
  the subscript $\sigma$ is to be omitted.

For $U=0$, the momentum distribution shows a jump at the Fermi momentum $k=k_\text{F}$. Under the influence of the quench the jump
\be
\Delta_n (t) = \lim_{k\rightarrow k_\text{F}^+}n_{k,\sigma}(t)-\lim_{k\rightarrow k_\text{F}^-}n_{k,\sigma}(t)
\ee
is reduced.

The momentum distribution of the system can be calculated by the Fourier transformation  
of the one-particle correlation function
\begin{align}
\label{eq:green}
G_\sigma(\vec{r},t) = \langle 0|\hat{c}_\sigma(\vec{r},t)^{\phantom\dagger}\hat{c}_\sigma(0,t)^\dagger|0\rangle
\end{align}
where the expectation value is taken with respect to the non-interacting Fermi sea $|0\rangle$.
Thus the time dependence of the operators $\hat{c}_\sigma$ and $\hat{c}_\sigma^\dagger$ is needed. 

To capture the time evolution of these operators we use the
the following ansatz \cite{uhr09}
\begin{align}
\hat{c}_\sigma^\dagger(\vec{r},t) = \hat{T}_{\vec{r}}^\dagger + \hat{T}_{\vec{r}}^\dagger \left(\hat{T}^\dagger\hat{L}^\dagger\right)_{\vec{r}} + ...
\label{ansatz}
\end{align}
with $\hat{T}^\dagger$ ($\hat{L}^\dagger$) denoting a general superposition of
 particle (hole) creation operators. For instance 
$\hat{T}^\dagger$ is given by
\begin{align}
\hat{T}_{\vec{r}}^\dagger = \sum_{|\vec{\delta}|\lessapprox v_{\text{max}} t} 
\sum_\sigma h_0(\vec{\delta}, t) \hat{c}_{\vec{r}+\vec{\delta},\sigma}^\dagger\,,
\label{eq:opT}
\end{align} 
where the creation operators acting on site $\vec{r}+\vec{\delta}$ are summed weighted with the prefactor $h_0(\vec{\delta},t)$. The shifts $\vec{\delta}$ are roughly bounded by the distance
over which quasi-particles can move in the time $t$ which is given by $v_{\text{max}}t$.
This is the light-cone effect \cite{calab06} first derived by Lieb and Robinson \cite{lieb72}. The effect of correlations 
beyond this bound is exponentially small.
The more complex terms $\hat{T}_{\vec{r}}^\dagger \left(\hat{T}^\dagger\hat{L}^\dagger\right)_{\vec{r}}$ 
are given by the superpositions of two creation and one annihilation operator. Another particle-hole pair 
$T^\dag L^\dag$ adds another creation and annihilation operator and so on.

The prefactors $h(\vec{\delta},t)$ contain the whole time dependence of the operators $\hat{T}_{\vec{r}}^\dagger$ and $\hat{L}_{\vec{r}}^\dagger$ 
and thus they determine the time dependence of $\hat{c}^\dagger(\vec{r},t)$.
To calculate the time dependence of the prefactors we use the equation of motion
\begin{align}
\label{eq:heisenberg}
\partial_t \hat{\text{A}}(\vec{r},t) = i\left[\hat{\text{H}},\hat{\text{A}}(\vec{r},t)\right]
\end{align}
 for the time derivative of any operator $\hat{A}$. 
 The advantage of dealing with the time dependence for the operators instead of the one
 for the quantum states is the dependence on the size of the system.
 For dealing with time dependent states we would be obliged to treat states
 in an infinite quantum system which is very difficult. In contrast, the appearance of the
 commutators in the time dependence of the operators implies a linked-cluster property.
 In other words, up to a certain order $m$ in the time $t$ one has to deal only with a
 finite number of operators while dealing with the infinite system in the thermodynamic limit.
 We stress that all results presented here refer to this limit.
  The caveat of the approach is that the number of operators grows exponentially with
 the order $m$.

When calculating the commutator $\left[\hat{\text{H}},\hat{c}^\dagger(\vec{r},t)\right]$ 
we encounter two cases. (i) The commutation with the non-interating 
part of the Hamiltonian $\hat{\text{H}}_0$ leads to a shift of single fermionic operators. 
(ii) The commutation with the interaction term $\hat{\text{H}}_{\text{int}}$
creates or annihilates additional particle-hole pairs $\left(\hat{T}^\dagger \hat{L}^\dagger\right)$. Iterating 
the commutation leads to the ansatz \eqref{ansatz} and extends it step by step.
Each commutation creates more and more terms with higher and higher number of particles and holes
involved.

Comparing the coefficients of the left hand side and of the right hand side of
the Heisenberg equation \eqref{eq:heisenberg} yields  differential equations for the prefactors $h$.
Then, this set of differential equations is solved numerically. The initial conditions 
of the prefactors are $h_0(0,0)=1$ and $h_i(\vec{r},t)=0 \; \forall i\neq 0$.

Since each commutation adds the coefficients necessary to describe another order in time $t$
the results become more and more accurate on increasing number of commutations (loops).
In this way a calculation with $n$ commutations provides results for $\hat{c}^\dagger(t)$ 
which are exact up to order $t^n$. 
To quantify the convergence of the results calculations with different numbers of commutations are performed
and compared. One may introduce a time $t_\text{runaway}$ up to which the deviations between the
results do not exceed a certain threshold, for instance $10^{-2}$. Thus, the precise definition of 
$t_\text{runaway}$ depends on the threshold, but for a given reasonable value of the
threshold one can clearly see that the results become more and more accurate for 
increasing number of loops. This has been performed for results for the Hubbard model
in Appendix A of Ref.\ \onlinecite{hamer13a} and results for the spinless
fermions can be found in  Appendix \ref{app:runaway} below.
The results in Appendix A of Ref.\ \onlinecite{hamer13a}
and those in Appendix \ref{app:runaway} below show that $t_\text{runaway}$ 
increases roughly quadratically with the number of loops $m$, i.e., $t_\text{runaway}\propto m^2$

\begin{figure}[htb]
    \begin{center}
    \includegraphics[width=0.95\columnwidth,clip]{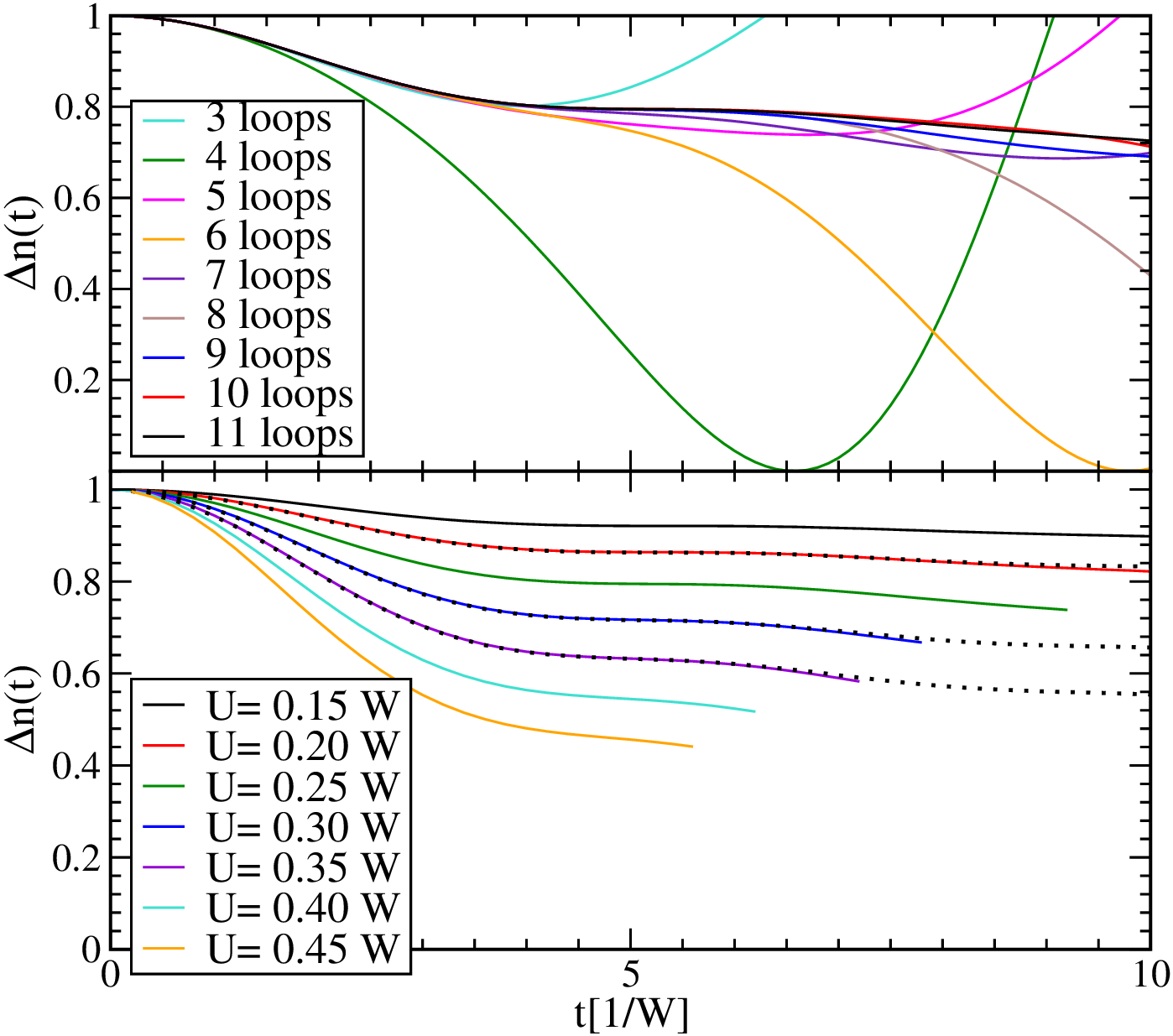}
    \end{center}
    \caption{(color online) Upper panel:
      Jump $\Delta n$(t) in the half filled spinless fermions model for various loop numbers. 
      Lower panel: 
      $\Delta n(t)$ for increasing $U$ (from top to bottom at small $t$) in 11 loops.
      The dotted lines display exemplary DMRG data from Karrasch et al.\ \cite{karra12b} illustrating the 
      	quantitative agreement in the range where our data is converged.
      \label{fig:varU}
}
\end{figure}

\subsection{Bosonization Results}
\label{ssec:boson}

We briefly recall what is to be expected for the jump in the momentum distribution
in Tomonaga-Luttinger models of non-interacting bosons \cite{cazal06,uhr09,iucci09,rentr12}.

For the spinless case one finds\cite{uhr09} 
\begin{equation}
\label{eq:powlaw-spinless}
\Delta n(t)= 
\Big[\frac{r^2}{r^2+(2v t)^2}\Big]^{2\gamma(1+\gamma)}
\end{equation}
where $v$ is the dressed velocity in the system and $r$ the characteristic length scale
of the interaction. The leading power law in time $t$ without the length scale $r$ 
in the denominator was first derived by Cazalilla \cite{cazal06}
 In the above formula, the interaction is assumed to range over all
momenta, i.e., from $-\infty$ to $\infty$. If a finite range in momentum space is assumed
oscillations occur. The occurrence of oscillations stemming from
the high-energy cutoff is indeed generic\cite{calab06,sabio10,rentr12}.
Note that such a finite range is natural in microscopic models
because  the finite extension of the Brillouin zone limits the range of the interaction
in momentum space.

The exponent in \eqref{eq:powlaw-spinless} is determined by $\gamma$ which is related to
the standard bosonization parameter in the usual way 
\begin{equation}
\gamma = (K+K^{-1}-2)/4.
\end{equation}
Note that the expression \eqref{eq:powlaw-spinless} is governed by 
an exponent related to the one occurring in equilibrium
\cite{meden92} except that $\gamma$ in equilibrium 
has been replaced by $2\gamma(1+\gamma)$ after the quench \cite{uhr09}. 
Thus, for small values
one finds a factor of 2, similar to the observation after quenches of other systems
\cite{moeck09a}. We stress that the replacement $\gamma \to 2\gamma(1+\gamma)$ 
is an inherent property of the Tomonaga-Luttinger model. It is not related
to any underlying microscopic model.

In the presence of spin, the above formulae are modified. The Hamiltonian
is given by the non-interacting sum of the spin part and of the charge part.
Consequently, the single-particle correlation \eqref{eq:green} and hence the
jump are given by the product of the responses in the spin and the charge channel
\cite{uhr09}
\begin{equation}
\label{eq:powlaw-hubbard}
\Delta n(t)= 
\Big[\frac{r_\rho^2}{r_\rho^2+(2v_\rho t)^2}\Big]^{\gamma_\rho(1+\gamma_\rho)}
\Big[\frac{r_\sigma^2}{r_\sigma^2+(2v_\sigma t)^2}\Big]^{\gamma_\sigma(1+\gamma_\sigma)}
\end{equation}
where $\nu\in\{\rho,\sigma\}$ stands for charge ($\rho$) or spin ($\sigma$) channel.
Again, the parameters $r_\nu$ are the characteristic length scales of the interaction, 
$v_\nu$ the velocities and $\gamma_\nu$ the equilibrium exponents.
The exponents can be expressed through the anomalous dimensions $K_\nu$ in the usual way
\begin{equation}
\gamma_\nu = (K_\nu+K_\nu^{-1}-2)/4.
\end{equation}
Note that the exponents in each channel separately take only half the value of their
counterpart in the spinless case. This is again the same as in equilibrium \cite{meden92,voit95}.
But also in the case with spin, the non-equilibrium exponents are obtained from
the equilibrium ones by the replacement $\gamma_\nu \to 2\gamma_\nu(1+\gamma_\nu)$.


\section{Results}

\subsection{Spinless Fermions}
\label{chap:results1}

In this section, we present the results obtained by the equation of motion approach
for the behavior of the spinless fermion model after the quench. 
We focus on short and intermediate 
times after the quench and show results for the jump $\Delta n(t)$. 
We explain the data from the equation of motion approach by comparing them
to the bosonization results.

\begin{figure}[ht]
    \begin{center}
    \includegraphics[width=0.95\columnwidth,clip]{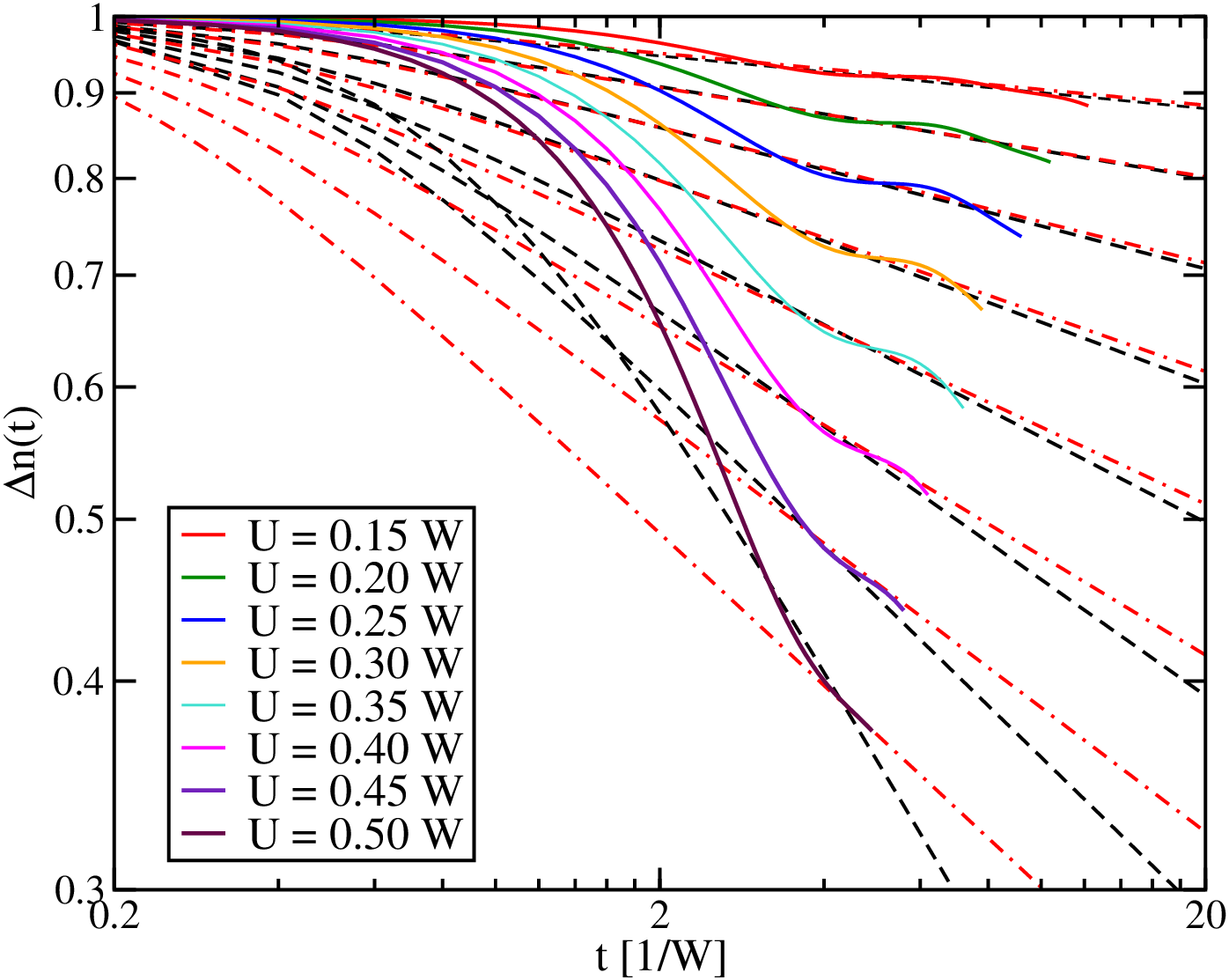}
    \end{center}
    \caption{(color online) 
      Solid lines: $\Delta n(t)$ in the half filled model of spinless fermions for increasing $U$ (from
      top to bottom) in 11 loops. Dashed (black) lines: $\Delta n(t)$ as in Eq.\ \eqref{eq:powlaw-spinless}
      with the ground state (GS) exponents from Eq.\ \eqref{eq:NN-Bethe}.
      Dashed-dotted (red) lines: 
      $\Delta n(t)$ with the Fermi sea (FS) exponents from Eq.\ \eqref{eq:NN-FS}.
      \label{fig:NNvarU}
}
\end{figure}

In Fig.\ \ref{fig:NNvarU} the jump $\Delta n(t)$ is shown for
the spinless fermion model with nearest-neighbor repulsion. 
The data agrees very well with the data by Karrasch et al.\ in Fig.\ 1 
of Ref.\ \onlinecite{karra12b} obtained by time dependent infinite-size DMRG,
for exemplary comparisons see Fig.\ \ref{fig:varU}.

\begin{figure}[ht]
    \begin{center}
    \includegraphics[width=0.95\columnwidth,clip]{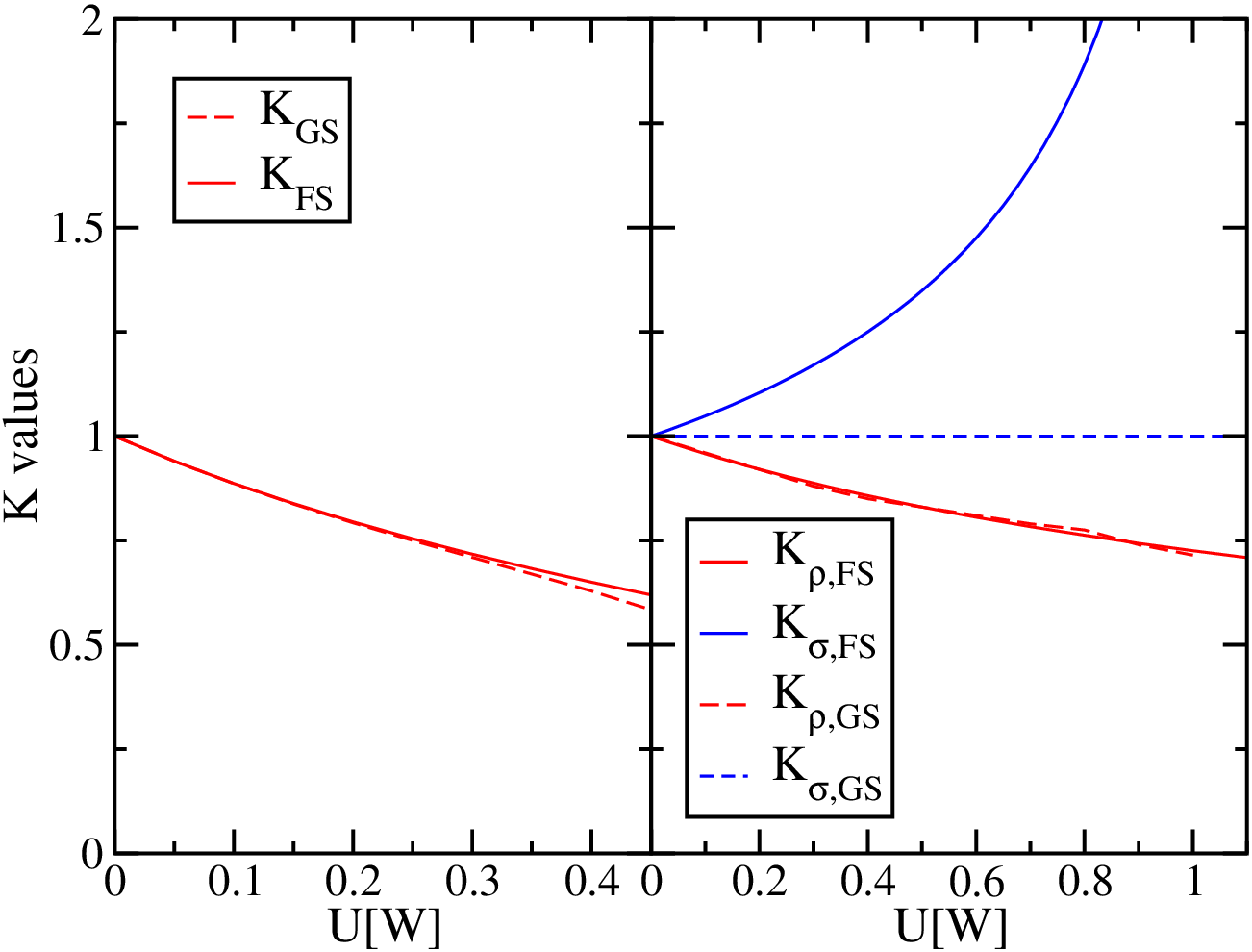}
    \end{center}
    \caption{(color online) 
   Results for the anomalous dimensions $K_\nu, \nu\in\{\rho,\sigma\}$ 
   	obtained by bosonization. Dashed lines: 
   Results obtained by bosonization
   around the ground state (GS) based on Bethe ansatz. 
   Solid lines: $K_\nu$ values obtained by bosonization around the Fermi sea (FS).
   Left panel: Results for the spinless fermion model with nearest-neighbor repulsion.
   Right panel: Results for the Hubbard chain at quarter-filling; the Bethe ansatz is
   difficult to evaluate so that slight inaccuracies imply some minor wiggling of the
   dashed curve for $K_\rho$.
      \label{fig:K_comp}
}
\end{figure}

First, we compare the data to the power law behavior \eqref{eq:powlaw-spinless}.
This still leaves the question which parameters are to be used.
Certainly, a first trial are the parameters which describe the models
in equilibrium. Since we are dealing with the systems at zero temperature
the `equilibrium' refers to the ground state and its immediate vicinity, i.e.,
the elementary excitations. Then, the anomalous dimension $K$ and the velocity can be 
determined by Bethe ansatz \cite{yang66a,luthe75}. They read
\begin{subequations}
\label{eq:NN-Bethe}
\begin{eqnarray}
K_\text{GS} &=& \pi/[2(\pi-\arccos(2U/W))]\\
v_\text{GS} &=& \pi\sin(\arccos(2U/W))/[2\arccos(2U/W)], \qquad
\end{eqnarray}
\end{subequations}
where we use the subscript GS for `ground state' to emphasize that these
parameters pertain to the behavior of the model at the lowest energies
in the vicinity of the ground state. The cutoff length $r$ of the curves
is fitted and evolves from $0.2$ to $0.6$ on increasing $U$, 
assuming that the lattice constant is set to unity.
The formulae \eqref{eq:NN-Bethe} are reasonable only up to $U=W/2$ where the
system enters a gapped phase.
The resulting anomalous dimension $K$ is depicted in the left panel of
Fig.\ \ref{fig:K_comp} as dashed curve.

Inspecting Fig.\ \ref{fig:NNvarU}  we see that the microscopic model
displays oscillations in $\Delta n(t)$ which are absent in Eq.\ \eqref{eq:powlaw-spinless}.
These oscillations are to be ascribed to the momentum cutoff of the interaction
in microscopic models \cite{calab06,sabio10,rentr12}. Otherwise, the power law 
\eqref{eq:powlaw-spinless}
nicely describes the accessible dynamics, at least for not too large values
of the interaction. 
This observation agrees with the one by Karrasch et al.\ in Ref.\ \onlinecite{karra12b}.

We point out, however, that the agreement deteriorates for larger values
of the interaction, namely for $U=0.45W$ and for $U=0.5W$. Thus we pose the question whether
the equilibrium exponent $\gamma$ is still the relevant one for non-equilibrium
situations. We stress that the answer to this question in an unconditioned `Yes'
for the Tomonaga-Luttinger model itself. But for any microscopic model or
for any model containing boson-boson interaction such as for instance the sine-Gordon model
the behavior in the vicinity of the ground state will in general be
\emph{different} from the behavior at higher energies. 
This means that if we stick to the description of the quench of a microscopic 
model in terms of an approximate field theoretic model it must be expected 
that the parameters of this approximate model depend on the initial conditions,
for example the extent of the quench. In the renormalization description 
of the sine-Gordon model off equilibrium by Mitra
and Giamarchi \cite{mitra11,mitra12} such a dependence of the anomalous dimension
$K$ on the initial condition has appeared explicitly,
even though the quench considered by them is not the same as here because
the sine term is switched on only adiabatically long time after the sudden quench.

In order to assess how far a quench puts the microscopic model
away from equilibrium, i.e., from the ground state, we define the
quench energy $\Delta E$
\be
\label{eq:ex_energy}
\Delta E := \langle \text{FS}| H(t>0) |\text{FS}\rangle - 
\langle \text{GS}| H(t>0) |\text{GS}\rangle
\ee
where $|\text{FS}\rangle$ stands for the initial state which in our case
is a Fermi sea while $|\text{GS}\rangle$ stands for the ground state of the
Hamiltonian after the quench. Thus, $\Delta E$ as defined above
measures the total excitation energy above the ground state induced by the quench.
In Ref.\ \onlinecite{grand10}, this quantity is called the heat.
It is conserved since the energy in a closed constant quantum system is 
a conserved quantity. We stress that quenches in \emph{imaginary}
time \cite{karra12b} which obey $|t\rangle=\exp(-H \tau)|\text{FS}\rangle$ 
do not conserve $\Delta E$ and thus behave differently.

\begin{figure}[htb]
    \begin{center}
    \includegraphics[width=0.95\columnwidth,clip]{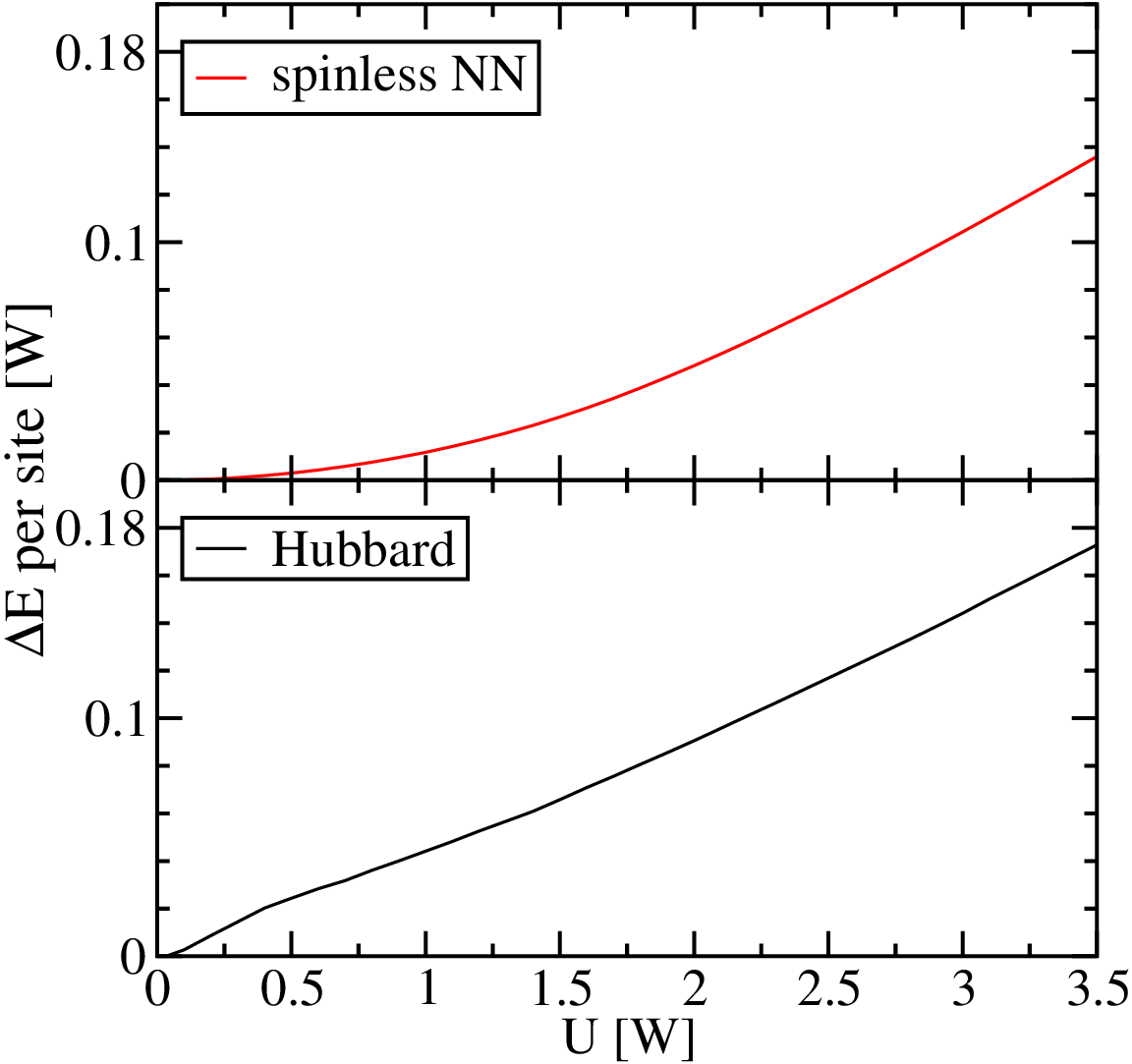}
    \end{center}
    \caption{(color online) 
 Excitation energies $\Delta E$ per site defined in \eqref{eq:ex_energy}
 in dependence on the interaction strength $U$ for the 
 spinless fermions model at half-filling (upper panel) and for the  
 quarter-filled Hubbard model (lower panel); the behavior at low values
 of $U$ is quadratic even though the difficult evaluation of the
 Bethe ansatz equations hindered us to reach very high precision.
 The results are very accurate at large $U$.
      \label{fig:ex_energy}
}
\end{figure}

In Fig.\ \ref{fig:ex_energy}, it is plotted for the two models under study
as function of the quenched interaction. For the spinless fermion model
we note the quenched energy remains fairly small $\lessapprox 3\cdot10^{-3}W$ 
for quenches below the phase transition at $U=W/2$ \cite{yang66aa,yang66a}. 
Thus one may not be surprised that the
ground state parameters \eqref{eq:NN-Bethe} yield good agreement.
In Ref.\ \onlinecite{karra12b}, it was concluded that the dynamics
of interaction quenches is universally governed by the equilibrium
exponents.

An alternative argument to reach the parameters for the
relevant field theoretic model after a quench is the following.
The dynamics after the quench starts from the initial state, here
$|\text{FS}\rangle$. The short time dynamics implies the iterated
gradual excitation of particle-hole pairs \cite{uhr09} so that
up to moderate times only a limited number of particle-hole pairs
need to be described. Thus bosonization of the density fluctuations
around the initial states $|\text{FS}\rangle$ is expected to yield
a valid description of the dynamics at short and intermediate times.

The bosonization in the vicinity of the Fermi sea (FS) corresponds to
the bosonization in leading order in $U$ because no feedback
effects due to the interaction need to be included. This leads us to the 
expressions \cite{luthe75,halda80,penc93,voit95,miran03}
\begin{subequations}
\label{eq:NN-FS}
\begin{eqnarray}
K_\text{FS} &=& \sqrt{(\pi v_\text{F} -U)/(\pi v_\text{F}+ 3U)}\\
v_\text{FS} &=& (1/\pi)\sqrt{(\pi v_\text{F}+ U)^2- 4U^2}.
\end{eqnarray}
\end{subequations}
The anomalous dimensions $K$ are shown in the left panel of Fig.\ \ref{fig:K_comp}
The implied power laws for the jump $\Delta n$ are included in Fig.\ \ref{fig:NNvarU}
as dashed-dotted curves. The fitted cutoff length $r$ varies from $0.2$ to 
$0.1$ on rising $U$.

Obviously, the difference between the curves from the GS and from the FS parameters
is small, cf.\ Fig.\ \ref{fig:NNvarU}. This can be understood easily in view of the 
small excitation energies $\Delta E$. Nevertheless, we point out that
for the largest interactions $U$ the FS curves fit better to the
numerical data than the GS curves. This underlines the relevance of the question 
which field theoretic model is the appropriate one off equilibrium.

We note that the bosonization in the vicinity of the Fermi sea leaves
out certain effects that will become important at even stronger
quenches, for instance the curvature
of the dispersion or the effect of higher order umklapp scattering.
(At leading order in $U$, no umklapp scattering occurs for spinless fermions.)
This restricts the validity of the FS formulae \eqref{eq:NN-FS} to values of $U$
below $\pi v_\text{F}\approx 1.57W$, where $K_\text{FS}$ and $v_\text{FS}$ vanish.


\subsection{1D Hubbard Model}
\label{chap:results2}

Quenches in a model \emph{with} spin are by themselves of great interest
because such models are much closer to what is realized in strongly
correlated systems such as Mott insulators. Moreover, they display
dynamical transitions at half-filling if quenched strongly in infinite dimensions
\cite{eck09,schiro10} and in one dimension \cite{hamer13a}.

In the present article, however, we focus on weaker quenches for
fillings off half-filling. The reason is that we want to focus on models which are 
quenched within the metallic phase. Thus we focus on quarter-filling
which also suppresses umklapp scattering, at least to leading order,
so that a quench into a Mott insulating phase is avoided.
Furthermore, the Hubbard model at quarter-filling displays two further
interesting features. 

The first is shown in the right panel of 
Fig.\ \ref{fig:K_comp} where the anomalous dimensions in the charge
and in the spin channel are depicted. The GS parameters are 
deduced from the ground state properties as obtained from Bethe ansatz
\cite{schul90a,essle05}. Note that $K_{\sigma,\text{GS}}=1$ by spin rotational symmetry \cite{giama04}.
The FS parameters are again those of the
bosonization around the Fermi sea amounting up to the bosonization \cite{penc93,voit95}
in leading order in the quenched interaction $U$ 
\begin{subequations}
\label{eq:FS-Hub}
\begin{eqnarray}
K_{\rho,\text{FS}} &=& \sqrt{2\pi v_\text{F}/(2\pi v_\text{F}+2U)}
\\
K_{\sigma,\text{FS}} &=& \sqrt{2\pi v_\text{F}/(2\pi v_\text{F}-2U)}
\\
 v_{\rho,\text{FS}} &=& v_\text{F}\sqrt{1+U/(\pi v_\text{F})}
 \\
 v_{\sigma,\text{FS}} &=& v_\text{F}\sqrt{1-U/(\pi v_\text{F})}.
\end{eqnarray}
\end{subequations}
The right panel of Fig.\ \ref{fig:K_comp} shows that the charge dimensions
agree surprisingly well in both caculations $K_{\rho,\text{GS}}\approx K_{\rho,\text{FS}}$.
In contrast, the spin dimension $K_{\sigma,\text{GS}}$ and $K_{\sigma,\text{FS}}$
differ significantly. It is well known from the analysis of the renormalization of the
underlying sine-Gordon model that $K_{\sigma,\text{GS}}$ converges to its final
value only for exponentially small energy scales \cite{giama04}.

The second interesting aspect is the fact that in the Hubbard model
a moderate interaction quench implies much larger excitation energies,
see lower panel of Fig.\ \ref{fig:ex_energy}.
For instance, a quench to $U=0.8W$ yields an excitation energy of about
$0.036W$. Thus, in the Hubbard model we can more
easily study the effects of a larger distance from the equilibrium situation.\\
The behavior of the system after strong quenches can be found in Ref. \onlinecite{hamer13a}.

\begin{figure}[ht]
    \begin{center}
    \includegraphics[width=0.95\columnwidth,clip]{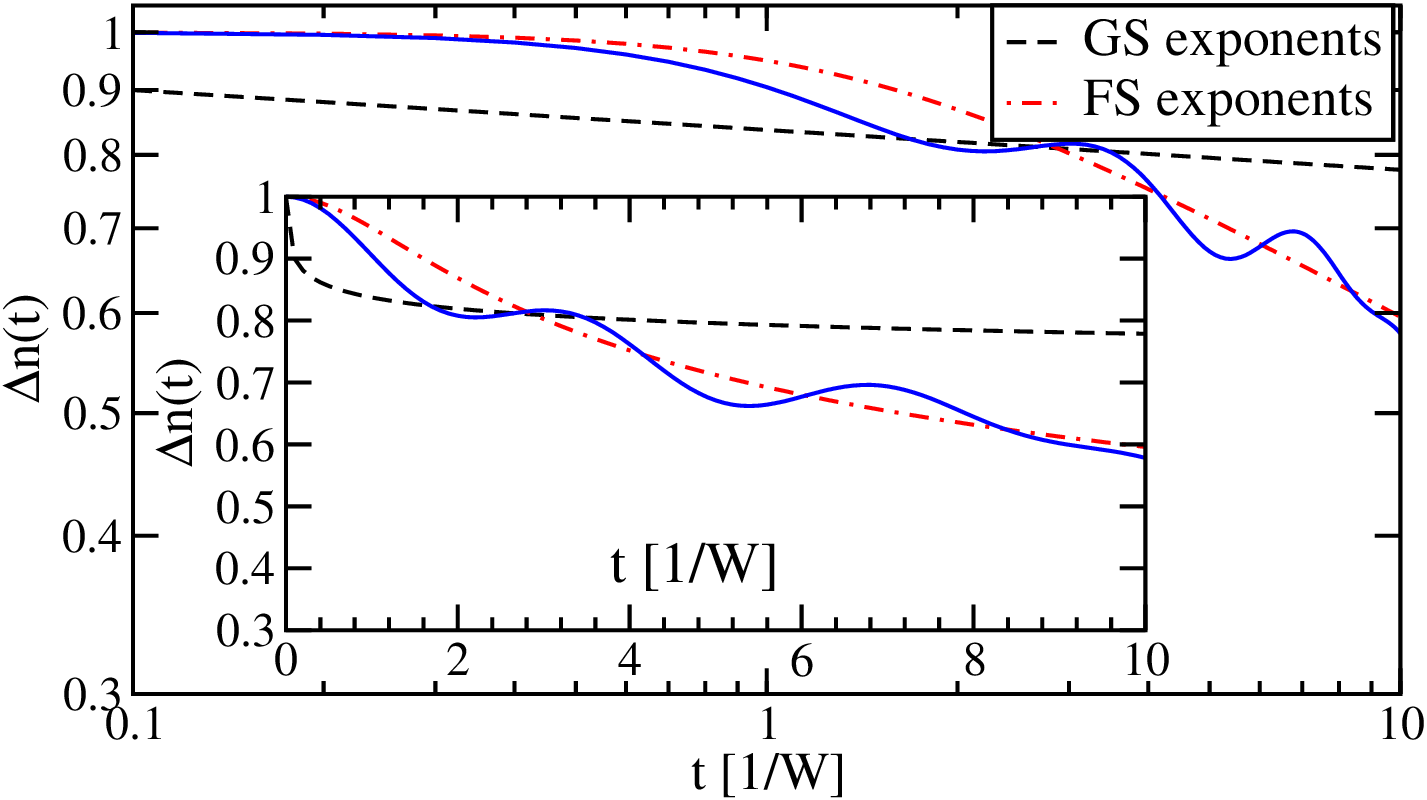}
    \end{center}
    \caption{(color online) 
       Jump $\Delta n(t)$ for the Hubbard model at quarter-filling at $U=0.8W$.  Lines as in Fig.\
       \protect\ref{fig:NNvarU}. For the GS exponents $r_\rho\approx0.01$ is fitted;
       $r_\sigma$ does not occur. 
       For the FS exponents $r_\rho\approx 1$ and $r_\sigma\approx 0.6$ are fitted.
      \label{fig:discrepan}
}
\end{figure}

Fig.\ \ref{fig:discrepan} shows the data for a quench to $U=0.8W$ in the
quarter-filled Hubbard model. The oscillations are again to be attributed
to the finite momentum cutoff of the interaction in a lattice model.

Inspecting the power laws, it is obvious that 
 the GS and the FS power laws differ significantly
as we expected from the significant differences in the anomalous dimensions
and the significant excitation energy of the quench.
We observe that the power law with the FS exponents fits much better than the 
power law with the GS exponents.
As can be seen in the inset of Fig.\ \ref{fig:discrepan}, 
the GS exponents can only be used to describe the curve over a small range
and the fitted range $r_\sigma$ is unreasonably small.

We recall that even at equilibrium  $K_\sigma$ reaches its final (GS) value only
for exponentially small energy scales $\varepsilon$. This implies that one must know the equal-time
Green function at two points in space with exponentially large distance $\vec{r}$
according to $\varepsilon\approx v_\text{F}/|\vec{r}|$ with the Fermi velocity
$v_\text{F}$. After the quench, the correlations due to the switched interaction
will develop gradually and spread out in time according to $|\vec{r}| \approx v_\text{max} t$,
i.e., $1/t$ acts as a low-energy cutoff.
Thus, effects at exponentially low energies are expected to show
up only at exponentially large times if they are not superseded by
other effects. For a discussion of the long-time behavior 
the reader is referred to the next section.

\begin{figure}[ht]
    \begin{center}
    \includegraphics[width=0.95\columnwidth,clip]{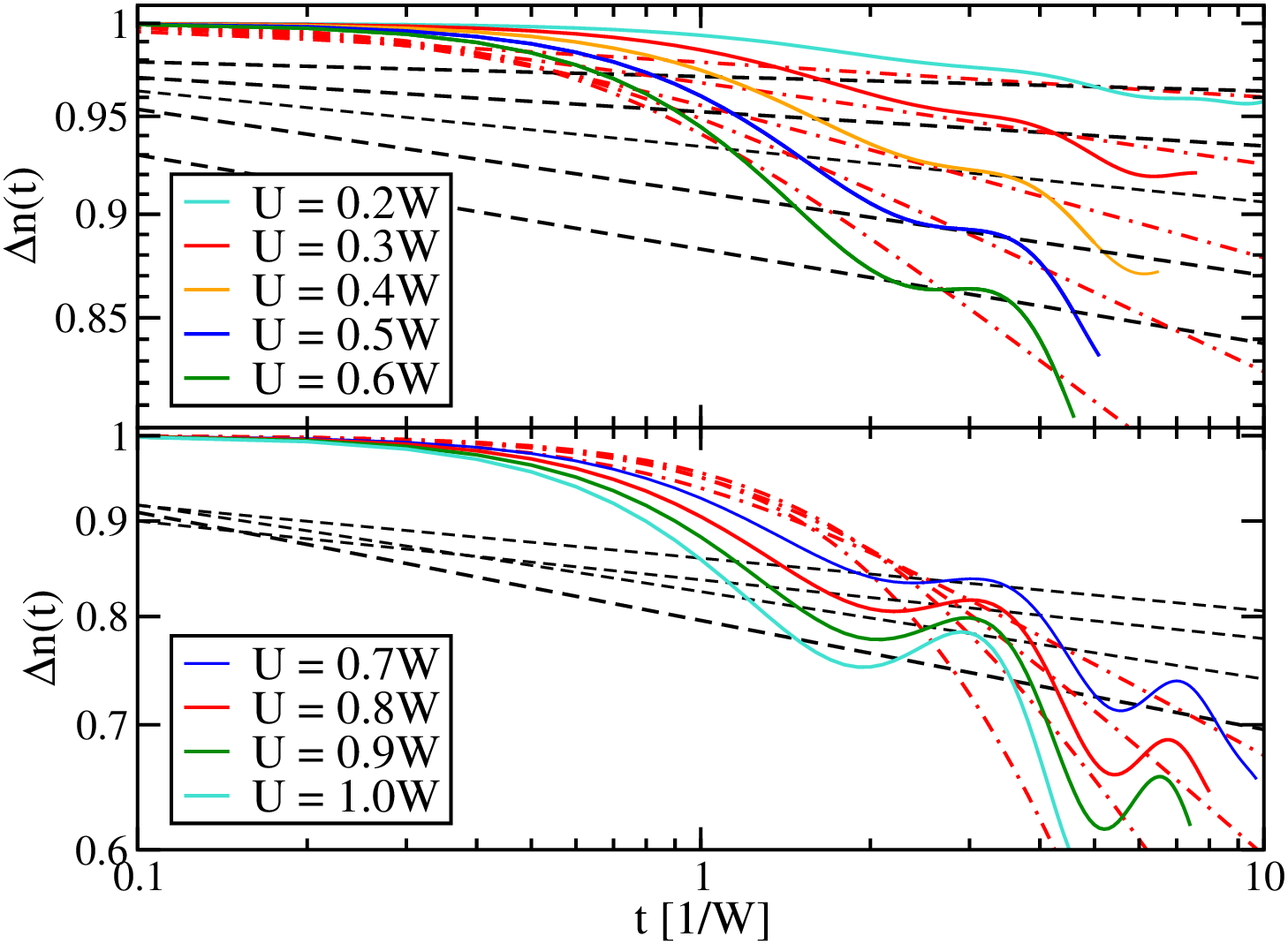}
    \end{center}
    \caption{(color online) 
   Solid lines: $\Delta n(t)$ for the Hubbard model at quarter filling for various values of $U$. 
   Dashed (black) lines: $\Delta n(t)$ given by the ground state (GS)
   exponents. Dashed-dotted (red) lines: 
   $\Delta n(t)$ calculated with the Fermi sea (FS) exponents.
      \label{fig:supp_hub}
}
\end{figure}

Here we state that the behavior at small and intermediate
times is described better by the FS parameters \eqref{eq:FS-Hub}.
To support this finding, Fig.\  \ref{fig:supp_hub} shows the jump $\Delta n(t)$ 
for the Hubbard model at quarter-filling for various interaction strengths $U$. 
As can be seen in the upper panel, for small values $U\lessapprox 0.3 W$ 
the differences between 
the GS and the FS power laws are small so that they are indistinguishable on the
accessible time scales.

For larger values of $U$, the time evolution of the jump is described very well
by the FS power law while the GS exponents do not fit the slope of the decay of the jump towards
larger times. Only the oscillations due to the high-energy cutoff\cite{calab06,sabio10,rentr12}
are missed by the power law \eqref{eq:powlaw-hubbard}.

For larger $U \gtrapprox W$ (lower panel) the agreement between microscopic
data and the FS power laws deteriorates.
We attribute this deterioration to the breakdown of the Tomonaga-Luttinger 
liquid description in terms of bosonic modes without interaction. 
Backscattering and the curvature of the single-particle dispersion
are neglected. Hence for large excitation energies $\Delta E$ which take 
the system even further way from its equilibrium these effects need
to be included.
For short and intermediate times, however, they are of reduced relevance.
For example, the leading correction due to backscattering is a term 
$\propto \int\cos(\sqrt{8}\Phi(x))dx$. But the expectation value of this and of any
of its higher derivatives vanishes with respect to the Fermi sea  because
in the latter the fluctuations $\langle\Phi(x)^2\rangle$ diverge which
smears out the cosine term completely to zero, see also the
discussion in Ref.\ \onlinecite{iucci10}.


\section{Behavior for Longer Times}
\label{chap:longtimes}

The next intriguing question is what happens for longer times $t$?
To date this question cannot be answered definitely. 
Numerically, no tools can treat \emph{long} times far off equilibrium.
Analytically, no general renormalization group theory
for nonequilibrium physics exists. A study of the sine-Gordon model on the basis
of Keldysh Green functions has been carried out \cite{mitra11,mitra12}. The non-interacting bosonic
system is quenched at $t=0$, but the sine term is switched on slowly afterwards.
At present, it is unclear to which extent this calculation captures the scenario
we are studying in the present article where the interaction is quenched abruptly
at $t=0$ in fermionic models.

To decide whether the results of the present article are relevant also for longer times 
we have to consider various energy cutoffs. The first was already mentioned above, namely the
inverse time $\propto 1/t$. Given the accessible times, see Figs.\ \ref{fig:varU}, 
\ref{fig:NNvarU}, and \ref{fig:supp_hub}, we find that this cutoff is of order of about
$0.1W$, the precise value depends on the value of the interaction, the filling and the model
under study. This is fairly large so that one may expect that the observed power laws 
will still gradually change for larger times.

But we come back to the distance of a quenched system to the true ground state which 
is measured by the excitation energy $\Delta E$ per site in \eqref{eq:ex_energy}, see Fig.\ 
\ref{fig:ex_energy}. We anticipate that this distance prevents the  parameters
of the effective low-energy model to reach their fixed-point values which
would define the relevant low-energy model just above the ground state.
At present, it is unclear whether $\Delta E/L$ itself sets the cutoff scale or whether
a fictitious temperature $T_\text{quench}$ which induces the same excitation energy
by thermal fluctuations, or yet another quantity sets the relevant scale.
The fictitious temperature $T_\text{quench}$ would be the most promising candidate if
the system relaxed towards a thermal state on moderate time scales. But so far all
numerical and analytical evidence in one-dimensional systems points against such a relaxation.
In particular, there is growing evidence that integrable models such as the sine-Gordon model
do no thermalize \cite{sabio10,fiore10,iucci10,grand10}. 
Their steady-state may be characterized by a generalized Gibbs ensemble \cite{rigol07}.

Mitra and Giamarchi \cite{mitra11,mitra12} argue that in interacting systems
the high-energy modes act as thermal bath to the low-energy modes so that the latter
behave  on long time scales as if they were thermally excited. A (small) dissipation 
rate is computed which sets the inverse time scale for the relaxation towards this 
apparent thermal behavior. Though this scenario appears intuitively plausible, 
the studied peculiar quench in the integrable sine-Gordon model 
leaves the question still open how generic it is.

In view of the above arguments, 
one may expect three temporal regimes in one-dimensional quenched systems: 
(i) Short time dynamics governed by power laws
with the FS exponents, (ii) Intermediate time dynamics governed by power laws
with slowly varying exponents which approach the GS values, but do not reach
them because their evolution is stopped by the distance of the system to
its equilibrium ground state. 
(iii) If the systems are not integrable, the dynamics of low-energy models
displays relaxation to thermal states at very long times.

Since at present, no scenario of the fascinating long time dynamics of one-dimensional 
systems, let alone of two- or three-dimensional systems, is firmly established further analytical and
numerical work is called for. In particular, the investigation of the potential
crossover regimes deserves further attention as well as the regime of very long times.


\section{Conclusions}

In the current work, we investigated the time evolution of the jump in the momentum distribution
of two one-dimensional fermionic models after interaction quenches. The method used was the
integration of the multiply iterated Heisenberg equations of motion for the operators (up to
about $10^7$ equations). This approach
circumvents the paramount difficulty to treat quantum states of infinite systems because
it focuses on the observables and their dynamics. In this dynamics, the commutation with
the local Hamilton operator implies a linked-cluster property \cite{uhr09} which keeps the problem
finite and tractable for each finite order of powers in the time $t$.

The first model studied comprises
spinless fermions with nearest-neighbor interaction at half-filling. 
For this model, a power law as found for the Tomonaga-Luttinger model provides 
a good description for the time evolution of the jump for short to intermediate times.
The exponents can be taken from the ground state properties known from Bethe ansatz
or from first order bosonization around the Fermi sea. They do not differ much as long
as the quenched systems is still metallic. The excitation energy induced
by the quench turns out to be remarkably small. The Fermi sea exponents fit slightly better
for the quenches close to the phase transition at $U=W/2$.

The second model studied is the Hubbard model at quarter filling for which much larger excitation
energies can be induced for $U$ values of the order of the band width $W$.
Also, the anomalous dimension $K_\sigma$ in the spin channel differs significantly depending on
the way it is determined, either around the Fermi sea or around the ground state.
On the accessible time scales, we found that the Fermi sea parameters describe
the temporal evolution better than the equilibrium parameters.

We discussed possible scenarios for longer times on the basis of energy cutoffs such as
the inverse time $1/t$, the excitation energy $\Delta E/L$, or a fictitious temperature
$T_\text{quench}$ leading thermally to the same excitation energy. The likely scenario 
is the occurrence of power laws with gradually changing exponents which start
from the Fermi sea values and approach the equilibrium values without reaching them due
to the distance of the quenched system to its equilibrium. For very long times, the 
low-energy modes may even show thermal relaxation if a macroscopic number of 
conserved quantities does not prevent it.
But many further studies are called for to clarify whether this likely scenario
is really the true one.


\begin{acknowledgments}
We are indebted to A.~Kl\"umper and F.~Essler for
help in the evaluation of the Bethe ansatz results, to V.~Meden for 
providing data, and to F.~Becca, S.~Kehrein, M.~Kollar, V.~Meden, 
M.~Moeckel, and J.~Stolze for inspiring discussions.
We acknowledge support from the Studienstiftung des
deutschen Volkes (SAH), from the Mercator Stiftung in project Pr-2011-003,
and from the DFG in project UH 90/5-1 (GSU).
\end{acknowledgments}


\begin{thebibliography}{54}
\expandafter\ifx\csname natexlab\endcsname\relax\def\natexlab#1{#1}\fi
\expandafter\ifx\csname bibnamefont\endcsname\relax
  \def\bibnamefont#1{#1}\fi
\expandafter\ifx\csname bibfnamefont\endcsname\relax
  \def\bibfnamefont#1{#1}\fi
\expandafter\ifx\csname citenamefont\endcsname\relax
  \def\citenamefont#1{#1}\fi
\expandafter\ifx\csname url\endcsname\relax
  \def\url#1{\texttt{#1}}\fi
\expandafter\ifx\csname urlprefix\endcsname\relax\def\urlprefix{URL }\fi
\providecommand{\bibinfo}[2]{#2}
\providecommand{\eprint}[2][]{\url{#2}}

\bibitem[{\citenamefont{Perfetti et~al.}(2006)\citenamefont{Perfetti, Loukakos,
  Lisowski, Bovensiepen, Berger, Biermann, Cornaglia, Georges, and
  Wolf}}]{perf06}
\bibinfo{author}{\bibfnamefont{L.}~\bibnamefont{Perfetti}},
  \bibinfo{author}{\bibfnamefont{P.~A.} \bibnamefont{Loukakos}},
  \bibinfo{author}{\bibfnamefont{M.}~\bibnamefont{Lisowski}},
  \bibinfo{author}{\bibfnamefont{U.}~\bibnamefont{Bovensiepen}},
  \bibinfo{author}{\bibfnamefont{H.}~\bibnamefont{Berger}},
  \bibinfo{author}{\bibfnamefont{S.}~\bibnamefont{Biermann}},
  \bibinfo{author}{\bibfnamefont{P.~S.} \bibnamefont{Cornaglia}},
  \bibinfo{author}{\bibfnamefont{A.}~\bibnamefont{Georges}}, \bibnamefont{and}
  \bibinfo{author}{\bibfnamefont{M.}~\bibnamefont{Wolf}},
  \bibinfo{journal}{Phys. Rev. Lett.} \textbf{\bibinfo{volume}{97}},
  \bibinfo{pages}{067402} (\bibinfo{year}{2006}).

\bibitem[{\citenamefont{Greiner
  et~al.}(2002{\natexlab{a}})\citenamefont{Greiner, Mandel, Esslinger,
  H\"ansch, and Bloch}}]{greiner02}
\bibinfo{author}{\bibfnamefont{M.}~\bibnamefont{Greiner}},
  \bibinfo{author}{\bibfnamefont{O.}~\bibnamefont{Mandel}},
  \bibinfo{author}{\bibfnamefont{T.}~\bibnamefont{Esslinger}},
  \bibinfo{author}{\bibfnamefont{T.~W.} \bibnamefont{H\"ansch}},
  \bibnamefont{and} \bibinfo{author}{\bibfnamefont{I.}~\bibnamefont{Bloch}},
  \bibinfo{journal}{Nature} \textbf{\bibinfo{volume}{415}}, \bibinfo{pages}{39}
  (\bibinfo{year}{2002}{\natexlab{a}}).

\bibitem[{\citenamefont{Greiner
  et~al.}(2002{\natexlab{b}})\citenamefont{Greiner, Mandel, H\"ansch, and
  Bloch}}]{greiner02b}
\bibinfo{author}{\bibfnamefont{M.}~\bibnamefont{Greiner}},
  \bibinfo{author}{\bibfnamefont{O.}~\bibnamefont{Mandel}},
  \bibinfo{author}{\bibfnamefont{T.~W.} \bibnamefont{H\"ansch}},
  \bibnamefont{and} \bibinfo{author}{\bibfnamefont{I.}~\bibnamefont{Bloch}},
  \bibinfo{journal}{Nature} \textbf{\bibinfo{volume}{419}}, \bibinfo{pages}{39}
  (\bibinfo{year}{2002}{\natexlab{b}}).

\bibitem[{\citenamefont{Schmidt and Monien}(2002)}]{schmi02}
\bibinfo{author}{\bibfnamefont{P.}~\bibnamefont{Schmidt}} \bibnamefont{and}
  \bibinfo{author}{\bibfnamefont{H.}~\bibnamefont{Monien}},
  \bibinfo{pages}{arXiv:/0202046v1}.

\bibitem[{\citenamefont{Freericks et~al.}(2006)\citenamefont{Freericks,
  Turkowski, and Zlati\'c}}]{freer06}
\bibinfo{author}{\bibfnamefont{J.~K.} \bibnamefont{Freericks}},
  \bibinfo{author}{\bibfnamefont{V.~M.} \bibnamefont{Turkowski}},
  \bibnamefont{and} \bibinfo{author}{\bibfnamefont{V.}~\bibnamefont{Zlati\'c}},
  \bibinfo{journal}{Phys. Rev. Lett.} \textbf{\bibinfo{volume}{97}},
  \bibinfo{pages}{266408} (\bibinfo{year}{2006}).

\bibitem[{\citenamefont{Eckstein et~al.}(2009)\citenamefont{Eckstein, Kollar,
  and Werner}}]{eck09}
\bibinfo{author}{\bibfnamefont{M.}~\bibnamefont{Eckstein}},
  \bibinfo{author}{\bibfnamefont{M.}~\bibnamefont{Kollar}}, \bibnamefont{and}
  \bibinfo{author}{\bibfnamefont{P.}~\bibnamefont{Werner}},
  \bibinfo{journal}{Phys. Rev. Lett.} \textbf{\bibinfo{volume}{103}},
  \bibinfo{pages}{056403} (\bibinfo{year}{2009}).

\bibitem[{\citenamefont{Daley et~al.}(2004)\citenamefont{Daley, Kollath,
  Schollw\"ock, and Vidal}}]{daley04}
\bibinfo{author}{\bibfnamefont{A.~J.} \bibnamefont{Daley}},
  \bibinfo{author}{\bibfnamefont{C.}~\bibnamefont{Kollath}},
  \bibinfo{author}{\bibfnamefont{U.}~\bibnamefont{Schollw\"ock}},
  \bibnamefont{and} \bibinfo{author}{\bibfnamefont{G.}~\bibnamefont{Vidal}},
  \bibinfo{journal}{J. Stat. Mech.} \textbf{\bibinfo{volume}{2004}},
  \bibinfo{pages}{P04005} (\bibinfo{year}{2004}).

\bibitem[{\citenamefont{White and Feiguin}(2004)}]{white04}
\bibinfo{author}{\bibfnamefont{S.~R.} \bibnamefont{White}} \bibnamefont{and}
  \bibinfo{author}{\bibfnamefont{A.~E.} \bibnamefont{Feiguin}},
  \bibinfo{journal}{Phys. Rev. Lett.} \textbf{\bibinfo{volume}{93}},
  \bibinfo{pages}{076401} (\bibinfo{year}{2004}).

\bibitem[{\citenamefont{Enss and Sirker}(2012)}]{enss12}
\bibinfo{author}{\bibfnamefont{T.}~\bibnamefont{Enss}} \bibnamefont{and}
  \bibinfo{author}{\bibfnamefont{J.}~\bibnamefont{Sirker}},
  \bibinfo{journal}{New J. Phys.} \textbf{\bibinfo{volume}{95}},
  \bibinfo{pages}{023008} (\bibinfo{year}{2012}).

\bibitem[{\citenamefont{Moeckel and Kehrein}(2008)}]{moe08}
\bibinfo{author}{\bibfnamefont{M.}~\bibnamefont{Moeckel}} \bibnamefont{and}
  \bibinfo{author}{\bibfnamefont{S.}~\bibnamefont{Kehrein}},
  \bibinfo{journal}{Phys. Rev. Lett.} \textbf{\bibinfo{volume}{100}},
  \bibinfo{pages}{175702} (\bibinfo{year}{2008}).

\bibitem[{\citenamefont{Schir\'o and Fabrizio}(2010)}]{schiro10}
\bibinfo{author}{\bibfnamefont{B.}~\bibnamefont{Schir\'o}} \bibnamefont{and}
  \bibinfo{author}{\bibfnamefont{M.}~\bibnamefont{Fabrizio}},
  \bibinfo{journal}{Phys. Rev. Lett.} \textbf{\bibinfo{volume}{105}},
  \bibinfo{pages}{076401} (\bibinfo{year}{2010}).

\bibitem[{\citenamefont{Sciolla and Biroli}(2011)}]{sciol11}
\bibinfo{author}{\bibfnamefont{B.}~\bibnamefont{Sciolla}} \bibnamefont{and}
  \bibinfo{author}{\bibfnamefont{G.}~\bibnamefont{Biroli}},
  \bibinfo{journal}{J. Stat. Mech.: Theor. Exp.} p. \bibinfo{pages}{P11003}
  (\bibinfo{year}{2011}).

\bibitem[{\citenamefont{Gambassi and Calabrese}(2011)}]{gamba11}
\bibinfo{author}{\bibfnamefont{A.}~\bibnamefont{Gambassi}} \bibnamefont{and}
  \bibinfo{author}{\bibfnamefont{P.}~\bibnamefont{Calabrese}},
  \bibinfo{journal}{Europhys. Lett.} \textbf{\bibinfo{volume}{95}},
  \bibinfo{pages}{66007} (\bibinfo{year}{2011}).

\bibitem[{\citenamefont{Mitra and Giamarchi}(2011)}]{mitra11}
\bibinfo{author}{\bibfnamefont{A.}~\bibnamefont{Mitra}} \bibnamefont{and}
  \bibinfo{author}{\bibfnamefont{T.}~\bibnamefont{Giamarchi}},
  \bibinfo{journal}{Phys. Rev. Lett.} \textbf{\bibinfo{volume}{107}},
  \bibinfo{pages}{150602} (\bibinfo{year}{2011}).

\bibitem[{\citenamefont{Goth and Assaad}(2012)}]{goth12}
\bibinfo{author}{\bibfnamefont{F.}~\bibnamefont{Goth}} \bibnamefont{and}
  \bibinfo{author}{\bibfnamefont{F.~F.} \bibnamefont{Assaad}},
  \bibinfo{journal}{Phys. Rev. B} \textbf{\bibinfo{volume}{85}},
  \bibinfo{pages}{085129} (\bibinfo{year}{2012}).

\bibitem[{\citenamefont{{Polkovnikov} et~al.}(2011)\citenamefont{{Polkovnikov},
  {Sengupta}, {Silva}, and {Vengalattore}}}]{polk11}
\bibinfo{author}{\bibfnamefont{A.}~\bibnamefont{{Polkovnikov}}},
  \bibinfo{author}{\bibfnamefont{K.}~\bibnamefont{{Sengupta}}},
  \bibinfo{author}{\bibfnamefont{A.}~\bibnamefont{{Silva}}}, \bibnamefont{and}
  \bibinfo{author}{\bibfnamefont{M.}~\bibnamefont{{Vengalattore}}},
  \bibinfo{journal}{Rev. Mod. Phys.} \textbf{\bibinfo{volume}{83}},
  \bibinfo{pages}{863} (\bibinfo{year}{2011}), \eprint{1007.5331}.

\bibitem[{\citenamefont{Barthel and Schollw\"ock}(2008)}]{barth08}
\bibinfo{author}{\bibfnamefont{T.}~\bibnamefont{Barthel}} \bibnamefont{and}
  \bibinfo{author}{\bibfnamefont{U.}~\bibnamefont{Schollw\"ock}},
  \bibinfo{journal}{Phys. Rev. Lett.} \textbf{\bibinfo{volume}{100}},
  \bibinfo{pages}{100601} (\bibinfo{year}{2008}).

\bibitem[{\citenamefont{Cazalilla}(2006)}]{cazal06}
\bibinfo{author}{\bibfnamefont{M.~A.} \bibnamefont{Cazalilla}},
  \bibinfo{journal}{Phys. Rev. Lett.} \textbf{\bibinfo{volume}{97}},
  \bibinfo{pages}{156403} (\bibinfo{year}{2006}).

\bibitem[{\citenamefont{Kollath et~al.}(2007)\citenamefont{Kollath, L\"auchli,
  and Altman}}]{kolla07}
\bibinfo{author}{\bibfnamefont{C.}~\bibnamefont{Kollath}},
  \bibinfo{author}{\bibfnamefont{A.~M.} \bibnamefont{L\"auchli}},
  \bibnamefont{and} \bibinfo{author}{\bibfnamefont{E.}~\bibnamefont{Altman}},
  \bibinfo{journal}{Phys. Rev. Lett.} \textbf{\bibinfo{volume}{98}},
  \bibinfo{pages}{180601} (\bibinfo{year}{2007}).

\bibitem[{\citenamefont{Manmana et~al.}(2007)\citenamefont{Manmana, Wessel,
  Noack, and Muramatsu}}]{manmana07}
\bibinfo{author}{\bibfnamefont{S.~R.} \bibnamefont{Manmana}},
  \bibinfo{author}{\bibfnamefont{S.}~\bibnamefont{Wessel}},
  \bibinfo{author}{\bibfnamefont{R.~M.} \bibnamefont{Noack}}, \bibnamefont{and}
  \bibinfo{author}{\bibfnamefont{A.}~\bibnamefont{Muramatsu}},
  \bibinfo{journal}{Phys. Rev. Lett.} \textbf{\bibinfo{volume}{98}},
  \bibinfo{pages}{210405} (\bibinfo{year}{2007}).

\bibitem[{\citenamefont{Uhrig}(2009)}]{uhr09}
\bibinfo{author}{\bibfnamefont{G.~S.} \bibnamefont{Uhrig}},
  \bibinfo{journal}{Phys. Rev. A} \textbf{\bibinfo{volume}{80}},
  \bibinfo{pages}{061602} (\bibinfo{year}{2009}).

\bibitem[{\citenamefont{D\'ora et~al.}(2011)\citenamefont{D\'ora, Haque, and
  Zar\'and}}]{dora11}
\bibinfo{author}{\bibfnamefont{B.}~\bibnamefont{D\'ora}},
  \bibinfo{author}{\bibfnamefont{M.}~\bibnamefont{Haque}}, \bibnamefont{and}
  \bibinfo{author}{\bibfnamefont{G.}~\bibnamefont{Zar\'and}},
  \bibinfo{journal}{Phys. Rev. Lett.} \textbf{\bibinfo{volume}{106}},
  \bibinfo{pages}{156406} (\bibinfo{year}{2011}).

\bibitem[{\citenamefont{Sabio and Kehrein}(2010)}]{sabio10}
\bibinfo{author}{\bibfnamefont{J.}~\bibnamefont{Sabio}} \bibnamefont{and}
  \bibinfo{author}{\bibfnamefont{S.}~\bibnamefont{Kehrein}},
  \bibinfo{journal}{New J. Phys.} \textbf{\bibinfo{volume}{12}},
  \bibinfo{pages}{055008} (\bibinfo{year}{2010}).

\bibitem[{\citenamefont{Fioretto and Mussardo}(2010)}]{fiore10}
\bibinfo{author}{\bibfnamefont{D.}~\bibnamefont{Fioretto}} \bibnamefont{and}
  \bibinfo{author}{\bibfnamefont{G.}~\bibnamefont{Mussardo}},
  \bibinfo{journal}{New J. Phys.} \textbf{\bibinfo{volume}{12}},
  \bibinfo{pages}{055015} (\bibinfo{year}{2010}).

\bibitem[{\citenamefont{Iucci and Cazalilla}(2010)}]{iucci10}
\bibinfo{author}{\bibfnamefont{A.}~\bibnamefont{Iucci}} \bibnamefont{and}
  \bibinfo{author}{\bibfnamefont{M.~A.} \bibnamefont{Cazalilla}},
  \bibinfo{journal}{New J. Phys.} \textbf{\bibinfo{volume}{12}},
  \bibinfo{pages}{055019} (\bibinfo{year}{2010}).

\bibitem[{\citenamefont{Grandi et~al.}(2010)\citenamefont{Grandi, Gritsev, and
  Polkovnikov}}]{grand10}
\bibinfo{author}{\bibfnamefont{C.}~\bibnamefont{De Grandi}},
  \bibinfo{author}{\bibfnamefont{V.}~\bibnamefont{Gritsev}}, \bibnamefont{and}
  \bibinfo{author}{\bibfnamefont{A.}~\bibnamefont{Polkovnikov}},
  \bibinfo{journal}{Phys. Rev. B} \textbf{\bibinfo{volume}{81}},
  \bibinfo{pages}{224301} (\bibinfo{year}{2010}).

\bibitem[{\citenamefont{Sciolla and Biroli}(2010)}]{sciol10}
\bibinfo{author}{\bibfnamefont{B.}~\bibnamefont{Sciolla}} \bibnamefont{and}
  \bibinfo{author}{\bibfnamefont{G.}~\bibnamefont{Biroli}},
  \bibinfo{journal}{Phys. Rev. Lett.} \textbf{\bibinfo{volume}{105}},
  \bibinfo{pages}{220401} (\bibinfo{year}{2010}).

\bibitem[{\citenamefont{Rigol et~al.}(2007)\citenamefont{Rigol, Dunjko,
  Yurovsky, and Olshanii}}]{rigol07}
\bibinfo{author}{\bibfnamefont{M.}~\bibnamefont{Rigol}},
  \bibinfo{author}{\bibfnamefont{V.}~\bibnamefont{Dunjko}},
  \bibinfo{author}{\bibfnamefont{V.}~\bibnamefont{Yurovsky}}, \bibnamefont{and}
  \bibinfo{author}{\bibfnamefont{M.}~\bibnamefont{Olshanii}},
  \bibinfo{journal}{Phys. Rev. Lett.} \textbf{\bibinfo{volume}{98}},
  \bibinfo{pages}{050405} (\bibinfo{year}{2007}).

\bibitem[{\citenamefont{Luther and Peschel}(1975)}]{luthe75}
\bibinfo{author}{\bibfnamefont{A.}~\bibnamefont{Luther}} \bibnamefont{and}
  \bibinfo{author}{\bibfnamefont{I.}~\bibnamefont{Peschel}},
  \bibinfo{journal}{Phys. Rev. B} \textbf{\bibinfo{volume}{12}},
  \bibinfo{pages}{3908} (\bibinfo{year}{1975}).

\bibitem[{\citenamefont{Haldane}(1980)}]{halda80}
\bibinfo{author}{\bibfnamefont{F.~D.~M.} \bibnamefont{Haldane}},
  \bibinfo{journal}{Phys. Rev. Lett.} \textbf{\bibinfo{volume}{45}},
  \bibinfo{pages}{1358} (\bibinfo{year}{1980}).

\bibitem[{\citenamefont{Meden and Sch\"onhammer}(1992)}]{meden92}
\bibinfo{author}{\bibfnamefont{V.}~\bibnamefont{Meden}} \bibnamefont{and}
  \bibinfo{author}{\bibfnamefont{K.}~\bibnamefont{Sch\"onhammer}},
  \bibinfo{journal}{Phys. Rev. B} \textbf{\bibinfo{volume}{46}},
  \bibinfo{pages}{15753} (\bibinfo{year}{1992}).

\bibitem[{\citenamefont{Penc and S\'olyom}(1993)}]{penc93}
\bibinfo{author}{\bibfnamefont{K.}~\bibnamefont{Penc}} \bibnamefont{and}
  \bibinfo{author}{\bibfnamefont{J.}~\bibnamefont{S\'olyom}},
  \bibinfo{journal}{Phys. Rev. B} \textbf{\bibinfo{volume}{47}},
  \bibinfo{pages}{6273} (\bibinfo{year}{1993}).

\bibitem[{\citenamefont{Voit}(1995)}]{voit95}
\bibinfo{author}{\bibfnamefont{J.}~\bibnamefont{Voit}}, \bibinfo{journal}{Rep.
  Prog. Phys.} \textbf{\bibinfo{volume}{58}}, \bibinfo{pages}{977}
  (\bibinfo{year}{1995}).

\bibitem[{\citenamefont{Miranda}(2003)}]{miran03}
\bibinfo{author}{\bibfnamefont{E.}~\bibnamefont{Miranda}},
  \bibinfo{journal}{Braz. J. Phys.} \textbf{\bibinfo{volume}{33}},
  \bibinfo{pages}{3} (\bibinfo{year}{2003}).

\bibitem[{\citenamefont{Giamarchi}(2004)}]{giama04}
\bibinfo{author}{\bibfnamefont{T.}~\bibnamefont{Giamarchi}},
  \emph{\bibinfo{title}{Quantum Physics in One Dimension}}
  (\bibinfo{publisher}{Clarendon Press, Oxford}, \bibinfo{year}{2004}).

\bibitem[{\citenamefont{Karrasch et~al.}(2012)\citenamefont{Karrasch, Rentrop,
  Schuricht, and Meden}}]{karra12b}
\bibinfo{author}{\bibfnamefont{C.}~\bibnamefont{Karrasch}},
  \bibinfo{author}{\bibfnamefont{J.}~\bibnamefont{Rentrop}},
  \bibinfo{author}{\bibfnamefont{D.}~\bibnamefont{Schuricht}},
  \bibnamefont{and} \bibinfo{author}{\bibfnamefont{V.}~\bibnamefont{Meden}},
  \bibinfo{journal}{Phys. Rev. Lett.} \textbf{\bibinfo{volume}{109}},
  \bibinfo{pages}{126406} (\bibinfo{year}{2012}).

\bibitem[{\citenamefont{Coira et~al.}(2012)\citenamefont{Coira, Becca, and
  Parola}}]{coira12}
\bibinfo{author}{\bibfnamefont{E.}~\bibnamefont{Coira}},
  \bibinfo{author}{\bibfnamefont{F.}~\bibnamefont{Becca}}, \bibnamefont{and}
  \bibinfo{author}{\bibfnamefont{A.}~\bibnamefont{Parola}}, p.
  \bibinfo{pages}{1205.2967} (\bibinfo{year}{2012}).

\bibitem[{\citenamefont{Pollmann et~al.}(2013)\citenamefont{Pollmann, Haque,
  and D\'ora}}]{pollm13}
\bibinfo{author}{\bibfnamefont{F.}~\bibnamefont{Pollmann}},
  \bibinfo{author}{\bibfnamefont{M.}~\bibnamefont{Haque}}, \bibnamefont{and}
  \bibinfo{author}{\bibfnamefont{B.}~\bibnamefont{D\'ora}},
  \bibinfo{journal}{Phys. Rev. B} \textbf{\bibinfo{volume}{87}},
  \bibinfo{pages}{041109(R)} (\bibinfo{year}{2013}).

\bibitem[{\citenamefont{Iucci and Cazalilla}(2009)}]{iucci09}
\bibinfo{author}{\bibfnamefont{A.}~\bibnamefont{Iucci}} \bibnamefont{and}
  \bibinfo{author}{\bibfnamefont{M.~A.} \bibnamefont{Cazalilla}},
  \bibinfo{journal}{Phys. Rev. A} \textbf{\bibinfo{volume}{80}},
  \bibinfo{pages}{063619} (\bibinfo{year}{2009}).

\bibitem[{\citenamefont{Rentrop et~al.}(2012)\citenamefont{Rentrop, Schuricht,
  and Meden}}]{rentr12}
\bibinfo{author}{\bibfnamefont{J.}~\bibnamefont{Rentrop}},
  \bibinfo{author}{\bibfnamefont{D.}~\bibnamefont{Schuricht}},
  \bibnamefont{and} \bibinfo{author}{\bibfnamefont{V.}~\bibnamefont{Meden}},
  \bibinfo{journal}{New J. Phys.} \textbf{\bibinfo{volume}{14}},
  \bibinfo{pages}{075001} (\bibinfo{year}{2012}).

\bibitem[{\citenamefont{Mitra and Giamarchi}(2012)}]{mitra12}
\bibinfo{author}{\bibfnamefont{A.}~\bibnamefont{Mitra}} \bibnamefont{and}
  \bibinfo{author}{\bibfnamefont{T.}~\bibnamefont{Giamarchi}},
  \bibinfo{journal}{Phys. Rev. B} \textbf{\bibinfo{volume}{85}},
  \bibinfo{pages}{075117} (\bibinfo{year}{2012}).

\bibitem[{\citenamefont{Hamerla and Uhrig}(2013)}]{hamer13a}
\bibinfo{author}{\bibfnamefont{S.~A.} \bibnamefont{Hamerla}} \bibnamefont{and}
  \bibinfo{author}{\bibfnamefont{G.~S.} \bibnamefont{Uhrig}},
  \bibinfo{journal}{Phys. Rev. B} \textbf{\bibinfo{volume}{87}},
  \bibinfo{pages}{064304} (\bibinfo{year}{2013}).

\bibitem[{43}]{note}
\bibnamefont{Note that version 1 of preprint 44 dealt with both 
subjects in a very short and not yet complete way so that we have decided to split and to
extend the discussion into the present article and Ref.\ 42.}

\bibitem[{\citenamefont{Hamerla and Uhrig}(2012)}]{hamer12}
\bibinfo{author}{\bibfnamefont{S.~A.} \bibnamefont{Hamerla}} \bibnamefont{and}
  \bibinfo{author}{\bibfnamefont{G.~S.} \bibnamefont{Uhrig}},
 \bibinfo{journal}{arXiv:/1207.2006v1}.

\bibitem[{\citenamefont{Fradkin}(1991)}]{fradk91}
\bibinfo{author}{\bibfnamefont{E.}~\bibnamefont{Fradkin}},
  \emph{\bibinfo{title}{Field Theories of Condensed Matter Systems}},
  vol.~\bibinfo{volume}{82} of \emph{\bibinfo{series}{Lecture Annote Series}}
  (\bibinfo{publisher}{Addison Wesley}, \bibinfo{address}{Redwood City},
  \bibinfo{year}{1991}).

\bibitem[{\citenamefont{Yang and Yang}(1966{\natexlab{b}})}]{yang66aa}
\bibinfo{author}{\bibfnamefont{C.~N.} \bibnamefont{Yang}} \bibnamefont{and}
  \bibinfo{author}{\bibfnamefont{C.~P.} \bibnamefont{Yang}},
  \bibinfo{journal}{Phys. Rev.} \textbf{\bibinfo{volume}{150}},
  \bibinfo{pages}{321} (\bibinfo{year}{1966}{\natexlab{b}}).

\bibitem[{\citenamefont{Yang and Yang}(1966{\natexlab{a}})}]{yang66a}
\bibinfo{author}{\bibfnamefont{C.~N.} \bibnamefont{Yang}} \bibnamefont{and}
  \bibinfo{author}{\bibfnamefont{C.~P.} \bibnamefont{Yang}},
  \bibinfo{journal}{Phys. Rev.} \textbf{\bibinfo{volume}{150}},
  \bibinfo{pages}{327} (\bibinfo{year}{1966}{\natexlab{a}}),
.

\bibitem[{\citenamefont{Hubbard}(1963)}]{hubba63}
\bibinfo{author}{\bibfnamefont{J.}~\bibnamefont{Hubbard}},
  \bibinfo{journal}{Phys. Roy. Soc. Lond.} \textbf{\bibinfo{volume}{276}},
  \bibinfo{pages}{238} (\bibinfo{year}{1963}).

\bibitem[{\citenamefont{Gutzwiller}(1963)}]{gutzw63}
\bibinfo{author}{\bibfnamefont{M.~C.} \bibnamefont{Gutzwiller}},
  \bibinfo{journal}{Phys. Rev. Lett.} \textbf{\bibinfo{volume}{10}},
  \bibinfo{pages}{159} (\bibinfo{year}{1963}).

\bibitem[{\citenamefont{Kanamori}(1963)}]{kanam63}
\bibinfo{author}{\bibfnamefont{J.}~\bibnamefont{Kanamori}},
  \bibinfo{journal}{Prog. Theor. Phys.} \textbf{\bibinfo{volume}{30}},
  \bibinfo{pages}{275} (\bibinfo{year}{1963}).

\bibitem[{\citenamefont{Lieb and Wu}(1968)}]{lieb68}
\bibinfo{author}{\bibfnamefont{E.~H.} \bibnamefont{Lieb}} \bibnamefont{and}
  \bibinfo{author}{\bibfnamefont{F.~Y.} \bibnamefont{Wu}},
  \bibinfo{journal}{Phys. Rev. Lett.} \textbf{\bibinfo{volume}{20}},
  \bibinfo{pages}{1445} (\bibinfo{year}{1968}).

\bibitem[{\citenamefont{Essler et~al.}(2005)\citenamefont{Essler, Frahm,
  Kl\"umper, and Korepin}}]{essle05}
\bibinfo{author}{\bibfnamefont{F.~H.~L.} \bibnamefont{Essler}},
  \bibinfo{author}{\bibfnamefont{H.}~\bibnamefont{Frahm}},
  \bibinfo{author}{\bibfnamefont{A.}~\bibnamefont{Kl\"umper}},
  \bibnamefont{and} \bibinfo{author}{\bibfnamefont{V.~E.}
  \bibnamefont{Korepin}} (\bibinfo{year}{2005}).

\bibitem[{\citenamefont{Calabrese and Cardy}(2006)}]{calab06}
\bibinfo{author}{\bibfnamefont{P.}~\bibnamefont{Calabrese}} \bibnamefont{and}
  \bibinfo{author}{\bibfnamefont{J.}~\bibnamefont{Cardy}},
  \bibinfo{journal}{Phys. Rev. Lett.} \textbf{\bibinfo{volume}{96}},
  \bibinfo{pages}{136801} (\bibinfo{year}{2006}).

\bibitem[{\citenamefont{Lieb and Robinson}(1972)}]{lieb72}
\bibinfo{author}{\bibfnamefont{E.~H.} \bibnamefont{Lieb}} \bibnamefont{and}
  \bibinfo{author}{\bibfnamefont{D.~W.} \bibnamefont{Robinson}},
  \bibinfo{journal}{Commun. math. Phys.} \textbf{\bibinfo{volume}{28}},
  \bibinfo{pages}{251} (\bibinfo{year}{1972}).

\bibitem[{\citenamefont{Moeckel and Kehrein}(2009)}]{moeck09a}
\bibinfo{author}{\bibfnamefont{M.}~\bibnamefont{Moeckel}} \bibnamefont{and}
  \bibinfo{author}{\bibfnamefont{S.}~\bibnamefont{Kehrein}},
  \bibinfo{journal}{Ann. of Phys.} \textbf{\bibinfo{volume}{324}},
  \bibinfo{pages}{2146} (\bibinfo{year}{2009}).

\bibitem[{\citenamefont{Schulz}(1990)}]{schul90a}
\bibinfo{author}{\bibfnamefont{H.~J.} \bibnamefont{Schulz}},
  \bibinfo{journal}{Phys. Rev. Lett.} \textbf{\bibinfo{volume}{64}},
  \bibinfo{pages}{2831} (\bibinfo{year}{1990}).

\end{thebibliography}

\appendix

\section{Convergence in the number of loops}
\label{app:runaway}

\begin{figure}[htb]
    \begin{center}
    \includegraphics[width=\columnwidth,clip]{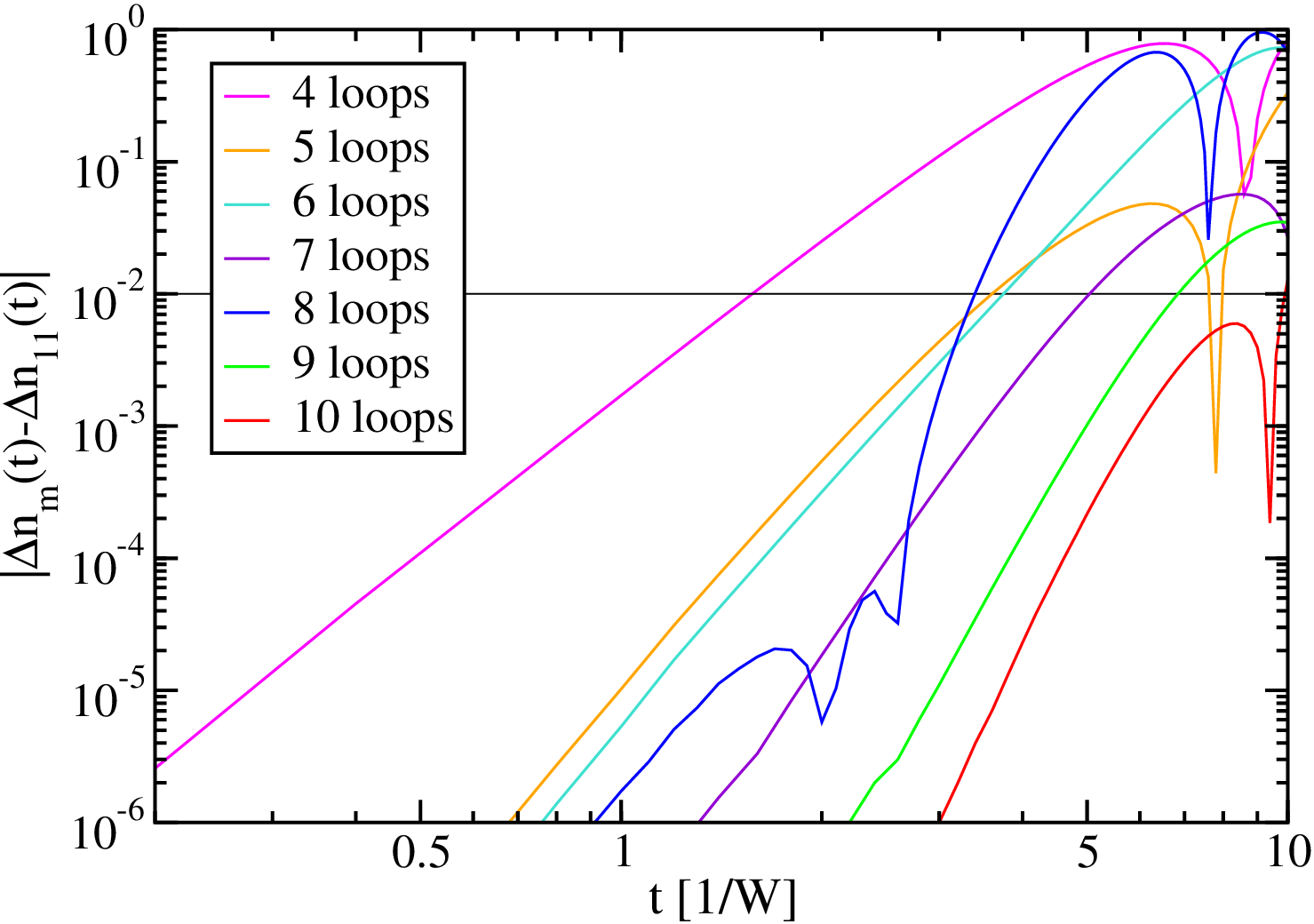}
    \end{center}
    \caption{(color online) 
  Absolute difference of the jump $\Delta n_m(t)$ at various numbers of loops $m$ 
  relative to the $11$-loop result $\Delta n_{11}(t)$   
  for the half filled spinless fermion model quenched by nearest-neighbor interaction. 
  Horizontal black line: Threshold for the determination of the runaway time.
      \label{fig:supp_conv1}
}
\end{figure}

\begin{figure}[htb]
    \begin{center}
    \includegraphics[width=\columnwidth,clip]{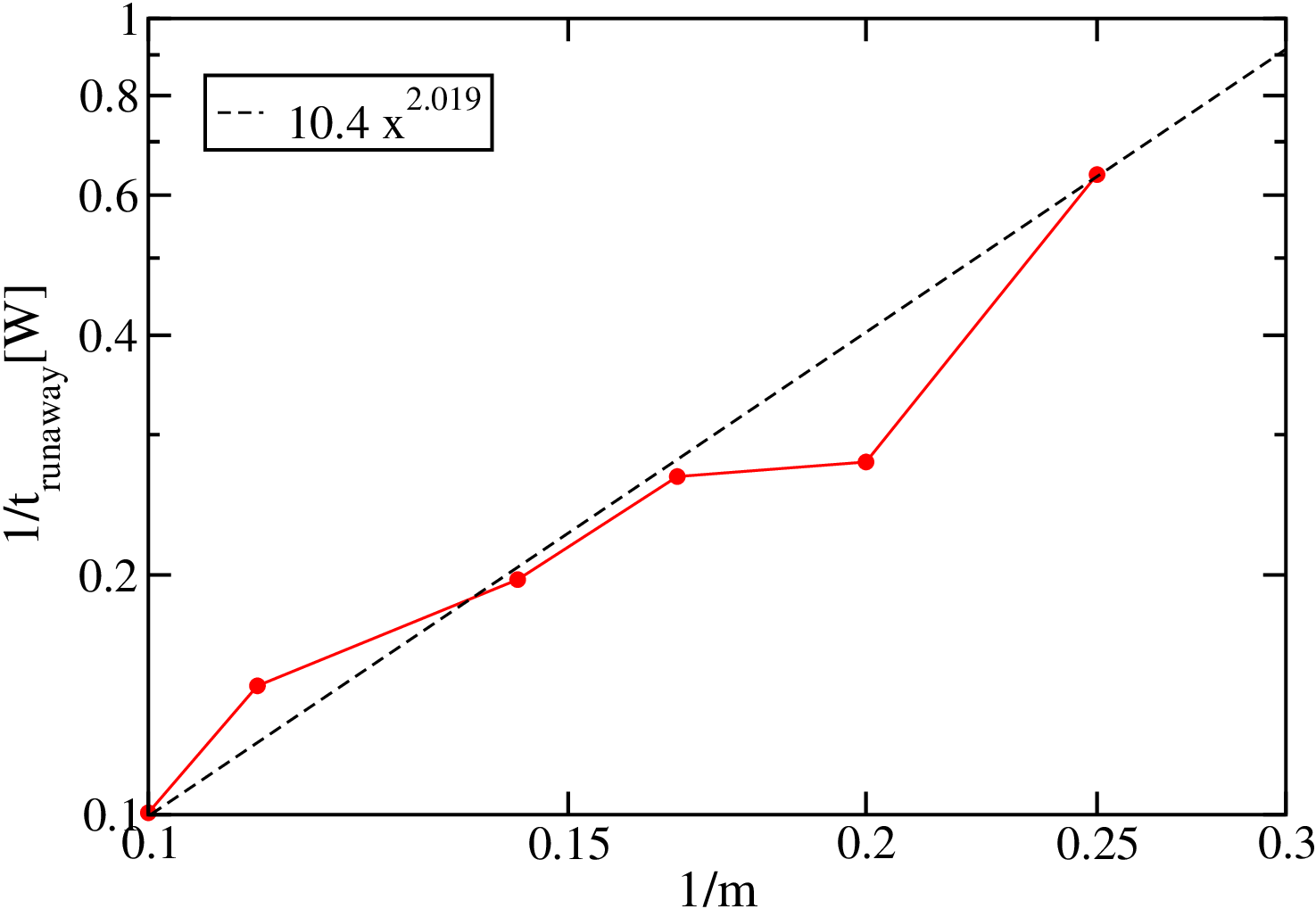}
    \end{center}
    \caption{(color online) 
   Double logarithmic plot of the inverse runaway time vs.\ the inverse number of loops of the corresponding calculation. Dashed line: Power law fit of the data with an exponent of about $2.02$.
      \label{fig:supp_conv2}
}
\end{figure}

In order to quantify the convergence properties of the approach, we
study the difference of the results for the jump in momentum distribution $\Delta n_m(t)$
obtained in $m$ loops. Taking the result with the highest number of loops (11) as reference,
Fig.\ \ref{fig:supp_conv1} depicts how the deviations from the 11-loop curve increase with time.
Setting a certain threshold for the deviation, here $0.01$, we determine up to which time
$t_\text{runaway}$ the deviation remains below the threshold. Fig.\ \ref{fig:supp_conv1}
shows data for the half filled spinless fermion case. Analogous data for the
Hubbard model can be found in Ref.\ \onlinecite{hamer13a}.

The choice of the threshold value is to some extent arbitrary, but it helps
to illustrate the main point: The results converge for increasing number of loops
$m\to\infty$. Finally, we plot the resulting inverse runaway times in the double logarithmic plot in
Fig.\ \ref{fig:supp_conv2} as function of $1/m$. Clearly, the times up to which the
results are reliable quickly increase on increasing loop number $m$. The exponent
is found to be of the order of $2$. Again, this is in agreement with
the previous findings for the Hubbard model \cite{hamer13a}.

\end{document}